\documentclass[%
 reprint,
superscriptaddress,
 amsmath,amssymb,
 aps,
]{revtex4-2}

\usepackage{graphicx}% Include figure files
\usepackage{dcolumn}% Align table columns on decimal point
\usepackage{bm}% bold math
\usepackage{physics}
\usepackage{upgreek}
\usepackage{gensymb}
\usepackage{units}
\usepackage{float}
\usepackage[colorlinks=true,linkcolor=blue,urlcolor=blue,citecolor=blue]{hyperref}

\usepackage{float}
\usepackage[commentmarkup=uwave,deletedmarkup=sout]{changes}
\definechangesauthor[name=Malte Grosche, color=blue]{fmg}
\begin{document}
\title{Direct Observation of the Spillover of High Magnetic Field-induced SC3 Superconductivity Outside the Spin-Polarized State in UTe$_2$}

\author{Zheyu Wu}
\author{Hanyi Chen}
\author{Theodore I. Weinberger}
\author{Mengmeng Long}
\affiliation{Cavendish Laboratory, University of Cambridge,\\
 JJ Thomson Avenue, Cambridge, CB3 0HE, United Kingdom}

\author{David Graf}
\affiliation{National High Magnetic Field Laboratory, Tallahassee, Florida, 32310, USA}

 \author{Andrej~Cabala}
 \author{Vladim\'{i}r~Sechovsk\'{y}}
 \author{Michal Vali{\v{s}}ka}
 \affiliation{Charles University, Faculty of Mathematics and Physics,\\ Department of
Condensed Matter Physics, Ke Karlovu 5, Prague 2, 121 16, Czech Republic}

\author{Gilbert G. Lonzarich}
\author{F. Malte Grosche}
\author{Alexander G. Eaton}
 \email{alex.eaton@phy.cam.ac.uk}
\affiliation{Cavendish Laboratory, University of Cambridge,\\
 JJ Thomson Avenue, Cambridge, CB3 0HE, United Kingdom}
 
\date{\today}

\begin{abstract}
\noindent
In our recent study of the high magnetic field phase landscape of UTe$_2$ [\href{https://doi.org/10.1103/PhysRevX.15.021019}{Phys. Rev. X \textbf{15}, 021019  (2025)}] we found indirect evidence that the SC3 superconducting phase spills out beyond the first-order phase boundary of the spin-polarized state. This prior study was limited to a maximal field strength of 41.5~T, and mapped the $b-ac$ rotation plane. Here we measure a high quality sample with residual resistivity ratio RRR~=~605 under rotations in the $b-c$ plane up to 45~T. This extended field range helps to unambiguously demonstrate the spillover of SC3 outside the polarized paramagnetic state. This is identified by the observation of zero resistance at low temperatures, for magnetic field strengths lower than the metamagnetic transition field resolved at higher temperatures. This observation is consistent with the scenario that electronic pairing of the SC3 phase is mediated by quantum critical fluctuations.

\end{abstract}

\maketitle 
\noindent
% The Pauli exclusion principle necessitates that rotationally symmetric superconductors, with an even-parity orbital component of the superconducting pair wavefunction, must have an antiparallel spin configuration~\cite{pauli1925,bauer2012book}. This is the case for the vast majority of known superconductors. By contrast, in a rare number of materials that lack rotational symmetry, pairs may combine with parallel spins and therefore must possess an odd-parity orbital wavefunction component~\cite{Balian-Werthamer1963,APMackenzieRevModPhys.75.657}. This additional degree of freedom in the spin structure can theoretically lead to rich superconducting phase diagrams comprising multiple proximate superconducting phases characterized by distinct order parameters~\cite{leggett_RevModPhys.47.331}.

\section{Introduction}

Heavy fermion UTe$_2$~\cite{Ran2019Science,Aoki_UTe2review2022,lewin2023review} is unique amongst known superconductors in possessing (at least) two magnetic field-induced superconducting states, which are distinct from the superconductivity found in zero field. Applying a magnetic field \textbf{H} along the hard magnetic $b$-axis acts to suppress the $T_c$ of the first (SC1) superconducting phase~\cite{tony2024enhanced}. Then, for $\mu_0 H \gtrapprox$~15~T the second (SC2) superconducting state emerges, persisting up to a critical field $\mu_0 H^* =$~34~T whereat the material crosses a first-order phase boundary into a spin-polarized magnetic state, quenching the superconductivity~\cite{Ranfieldboostednatphys2019,Knebel2019}. Tracking the metamagnetic phase boundary upon tilting \textbf{H} away from $b$ toward $c$ at an angle $\theta$ induces the third (SC3) superconductive state~\cite{Ranfieldboostednatphys2019}. Extensive high-$H$ measurements by several groups have discerned that at low temperature $T \approx$~0.5~K this pocket of superconductivity occupies a narrow angular domain between 19$\degree \lessapprox \theta \lessapprox 45\degree$, persisting to very high $\mu_0H >$~70~T~\cite{Ranfieldboostednatphys2019,helm2024,frank2024orphan,tony2025brief}. In addition to numerous contacted and contactless high-$H$ resistivity measurements, strong evidence that SC3 constitutes an intrinsic bulk superconducting state stems from the observation of vanishing Hall resistance~\cite{helm2024}, the presence of sudden adiabaticity upon crossing the SC3 phase boundary as resolved by magnetocaloric effect measurements~\cite{LANL_bulk_UTe2}, alongside ultrasound signatures characteristic of a vortex lattice~\cite{marquardt2025bulksignaturesreentrantsuperconductivity}.

Initial experimental studies of SC3 resolved this anomalously magnetophilic superconductivity exclusively within the spin-polarized state~\cite{lewin2023review}. Subsequent measurements, at magnetic field tilt angles away from the $b-c$ plane toward the $a$-axis, observed a toroidal domain of SC3 in three-dimensional magnetic field space, but still seemingly constrained within the spin-polarized state~\cite{lewin2025halo}. If SC3 were to exclusively reside within the spin-polarized host state, that could indicate the presence of a field-compensation mechanism underpinning the superconductivity~\cite{JaccPet1962PRL}.

Recently, we mapped the high field phase landscape of UTe$_2$~\cite{qcl} at various inclinations of \textbf{H}, to see how the menagerie of emergent electronic phases evolve throughout three-dimensional field space. We utilized several pulsed and steady magnet systems including an all-resistive 41.5~T steady magnet. While pulsed field systems provide higher maximal field values -- which are valuable due to the high field scales of several phenomena in UTe$_2$ -- steady magnet systems offer numerous advantages in that they can enable far more sensitive measurements, of intrinsic properties of the equilibrium state, at lower temperatures. Measuring the magnetoresistance upon rotating \textbf{H} from [010] toward [101] at 0.4~K, we observed a significant portion of the SC3 phase -- as identified by the observation of zero resistance -- that appeared to condense at $H < H^*$~\cite{qcl}. However, this experiment was limited to $\mu_0H \leq41.5$~T, and upon warming to $T>T_c$ the peak in magnetoresistance that characterizes $H^*$ was not observable below the critical endpoint temperature of the metamagnetic phase boundary. This therefore provided strong, albeit somewhat indirect, evidence that a portion of SC3 spills out below the metamagnetic phase boundary.

Here we present magnetotransport measurements of UTe$_2$ upon rotating \textbf{H} in the $b-c$ rotation plane for $\mu_0H \leq45$~T. Again we observe signatures of the zero resistance SC3 state forming at $H < H^*$ -- however, here we unambiguously show that SC3 occupies a small region of phase space outside the polarized paramagnetic state. This result has important implications regarding the likely physical mechanisms underpinning the formation of SC3 superconductivity.

\begin{figure}[t!]
    \includegraphics[width=\linewidth]{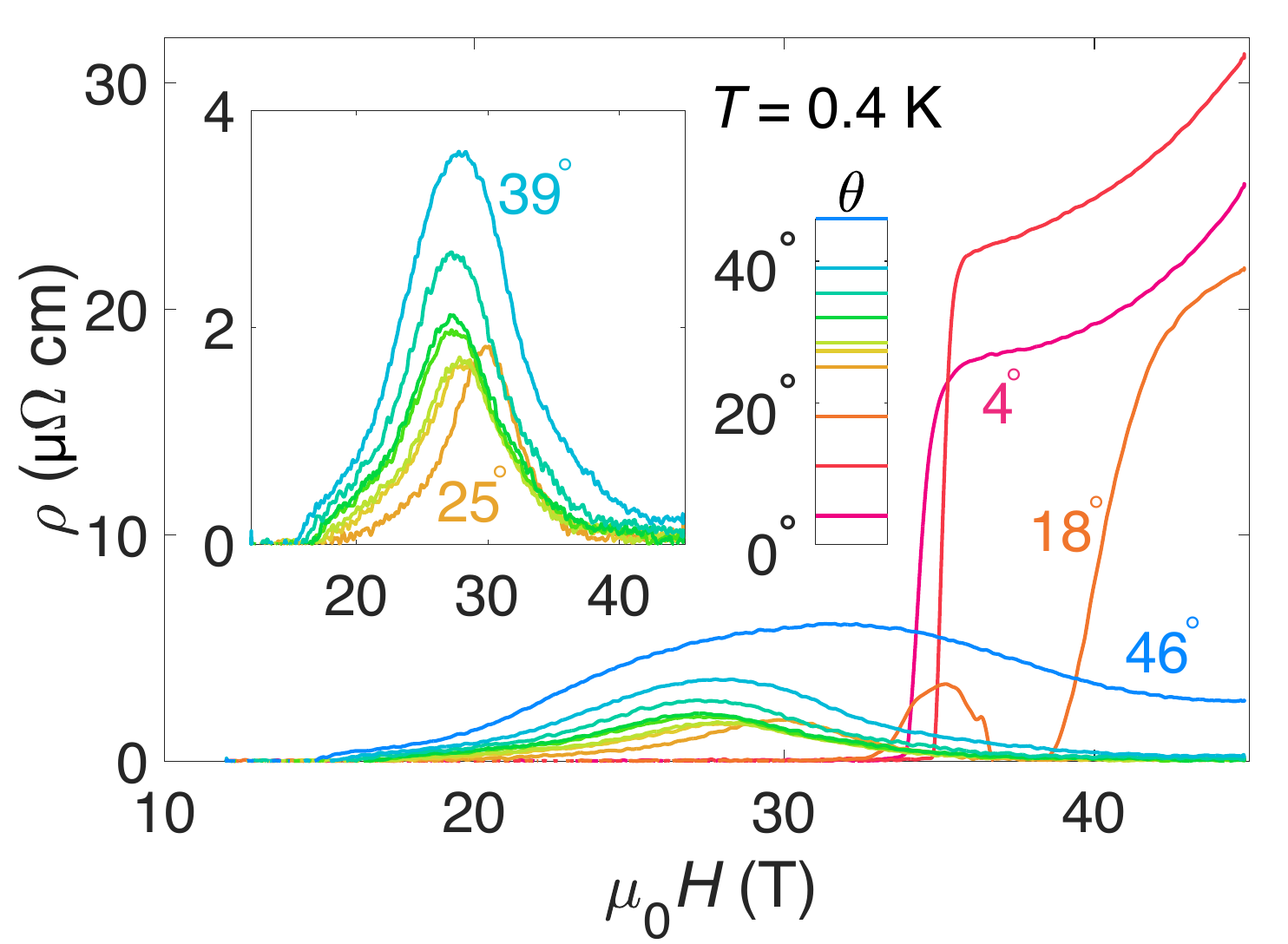}
    \caption{Resistivity $\rho$ versus magnetic field strength $H$ for rotations by angle $\theta$ in the $b-c$ plane. 0$\degree$ corresponds to field oriented along the $b$-axis, while 90$\degree$ represents the $c$-axis. The inset plots a zoom-in for a subset of curves in the range 25$\degree \leq \theta \leq 39\degree$, which exhibit an anomalous maximum in $\rho(H)$. We take this maximum to indicate the onset of the SC3 state.}
    \label{fig:angles}
\end{figure}

\section{Experimental methods}

The single crystal UTe$_2$ specimen investigated in this study was grown in a salt flux by the procedure specified in \cite{Eaton2024}. The sample was oriented by Laue diffractometry. 25~$\upmu$m gold wires were spot-welded onto the (001) surface to form electrical contacts. Magnetotransport data were obtained by applying current along the [100] direction using a Keithley 6221 current source at low frequency $<$50~Hz, with the resulting potential difference measured by an SR86x lock-in amplifier. All data were acquired with a 1~mA excitation amplitude, which translates to a current density of 0.45~$\mathrm{A cm^{-2}}$. Measurements were performed in the 45~T hybrid magnet at the National High Magnetic Field Laboratory, Tallahassee, Florida. This system combines an inner resistive solenoid, which can generate a field strength of up to 33.2~T, sitting within a large superconducting outsert coil at a steady 11.8~T. In combination this setup provides a maximal magnetic flux density of 45~T at the sample space. The minimum field strength of all the measurements presented here is, therefore, 11.8~T. The sample was mounted on a rotator probe, allowing in-situ rotation of the orientation of \textbf{H}, with angles calibrated by a Hall sensor. A $^3$He sorption cryostat was utilized, providing a base temperature of 0.4~K.

\section{Results}

\begin{figure}[t!]
    \includegraphics[width=\linewidth]{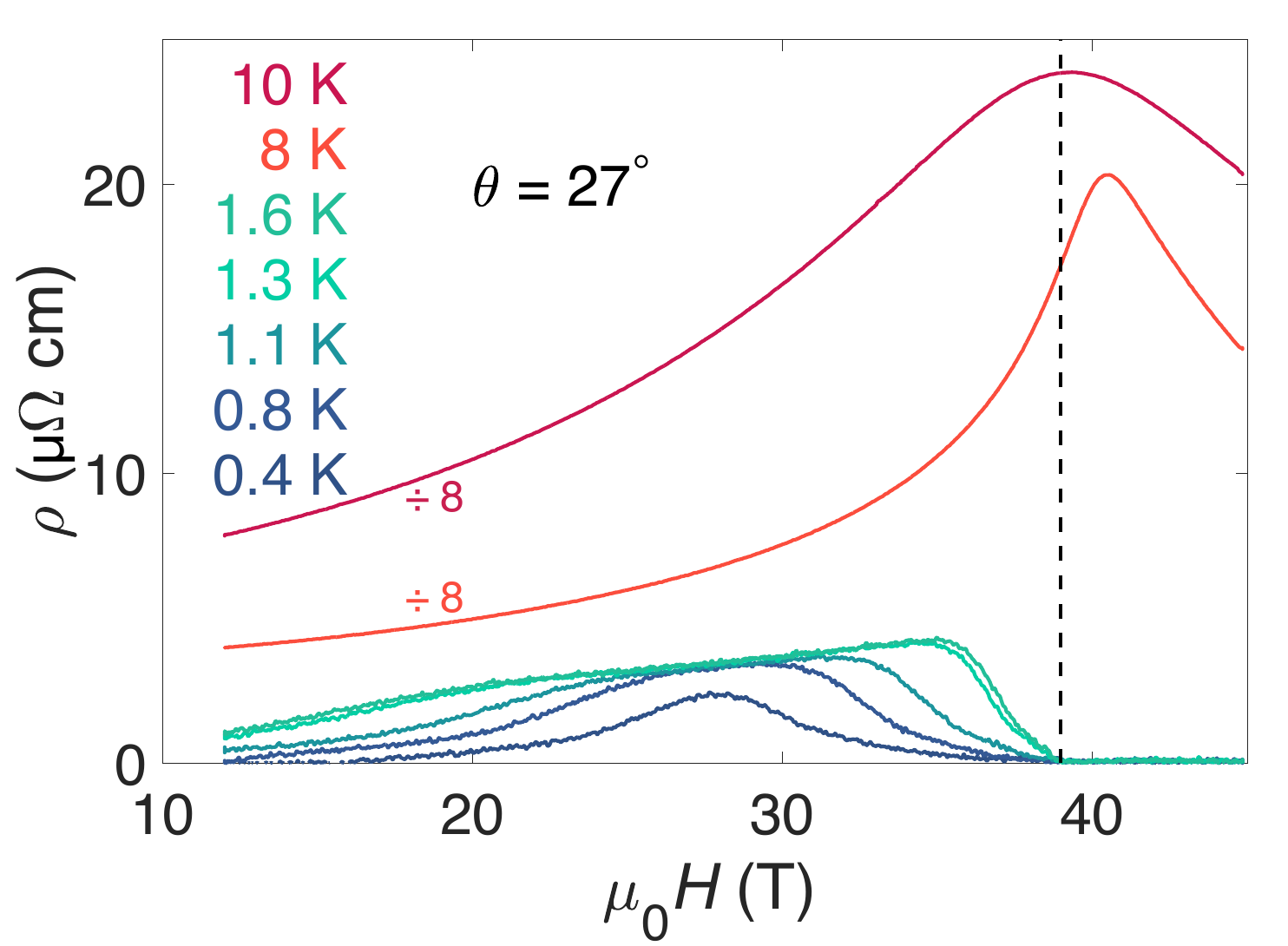}
    \caption{Temperature evolution of $\rho(H)$ with \textbf{H} tilted to $\theta = 25\degree$ for 0.4~K $\leq T \leq$ 10~K. The 8~K and 10~K curves have been rescaled by a factor of $\nicefrac{1}{8}$ for ease of comparison. The vertical dashed line marks the value of $H$ where zero resistivity at low $T$ identifies the low-$H$ boundary of the SC3 state. At 8~K the sharp peak in $\rho(H)$ -- identifying the metamagnetic transition into the spin-polarized state at $H^*$ -- clearly occurs at a higher value of $H$ than the dashed line.}
    \label{fig:temps}
\end{figure}

Fig.~\ref{fig:angles} plots the magnetoresistivity $\rho(H)$ of UTe$_2$ for rotations of \textbf{H} from \textbf{H} $\parallel b$ ($\theta =0\degree$) toward \textbf{H} $\parallel c$ ($\theta =90\degree$).  At $\theta = 4\degree$ zero resistance is observed up to 34~T, where $H^*$ is located, at which point $\rho$ jumps suddenly upon accessing the spin-polarized state. A positive slope of $\nicefrac{\partial \rho}{\partial H}$ is then observed up to 45~T. For $\theta =$ 18$\degree$ all three superconducting phases are accessed, with a small pocket of the normal state observed over the interval 32.3~T $\leq \mu_0H \leq$ 36.7~T. For higher angles of inclination, SC2 is no longer accessible for $\theta \geq$~25$\degree$, with the SC1-normal state transition observed at $\mu_0H <$~20~T. Interestingly, although $\nicefrac{\partial \rho}{\partial H}$ is initially positive after the transition for these higher angles, a maximum in $\rho(H)$ is then observed, followed by negative $\nicefrac{\partial \rho}{\partial H}$ until the SC3 state is accessed when $\rho$ becomes zero again (see inset to Fig.~\ref{fig:angles}). The minimal field value we observe $\rho = 0$ for SC3 is 36.0~T at $\theta=25\degree$.

In Fig.~\ref{fig:temps} we plot the temperature evolution of $\rho(H)$ for $\theta = 25\degree$. At $T = 0.4$~K the resistive transition between SC1 and the normal state is located at 17.2~T. $\rho(H)$ then rises gradually for increasing $H$ up to 28.2~T, at which point it reaches a maximum before turning over to give negative $\nicefrac{\partial \rho}{\partial H}$ until returning to zero at 38.0~T as SC3 is accessed. We measured four further temperatures in $^3$He liquid up to 1.6~K, before progressing to a gaseous sample environment to measure at 8~K and 10~K. At 8~K the metamagnetic transition is clearly resolved by a sharp peak in $\rho(H)$ \cite{knafo2019magnetic} located at 40.5~T. Prior studies have shown that for \textbf{H} in the $bc$ plane the metamagnetic transition shows very little variation in temperature below its critical endpoint (generally observed at $T \lessapprox$~9~K), with $H^*$ monotonically decreasing at elevated temperatures above the critical endpoint~\cite{Miyake2019,knafo2019magnetic,Miyake2021,valiska2024dramatic,qcl}, as we see here at 10~K. The location of $H^*$ we resolve at 8~K is thus a lower bound of $H^*$ in the zero temperature limit for this orientation of \textbf{H}. Therefore, given that zero resistance is observed at 38.0~T at low temperature whereas, above $T_c$, $\mu_0H^*$ is identified at 40.5~T, this directly shows that there is a narrow portion of the SC3 pocket of the UTe$_2$ phase diagram that spills out to occupy a region of $H < H^*$.

\begin{figure}[t!]
    \includegraphics[width=\linewidth]{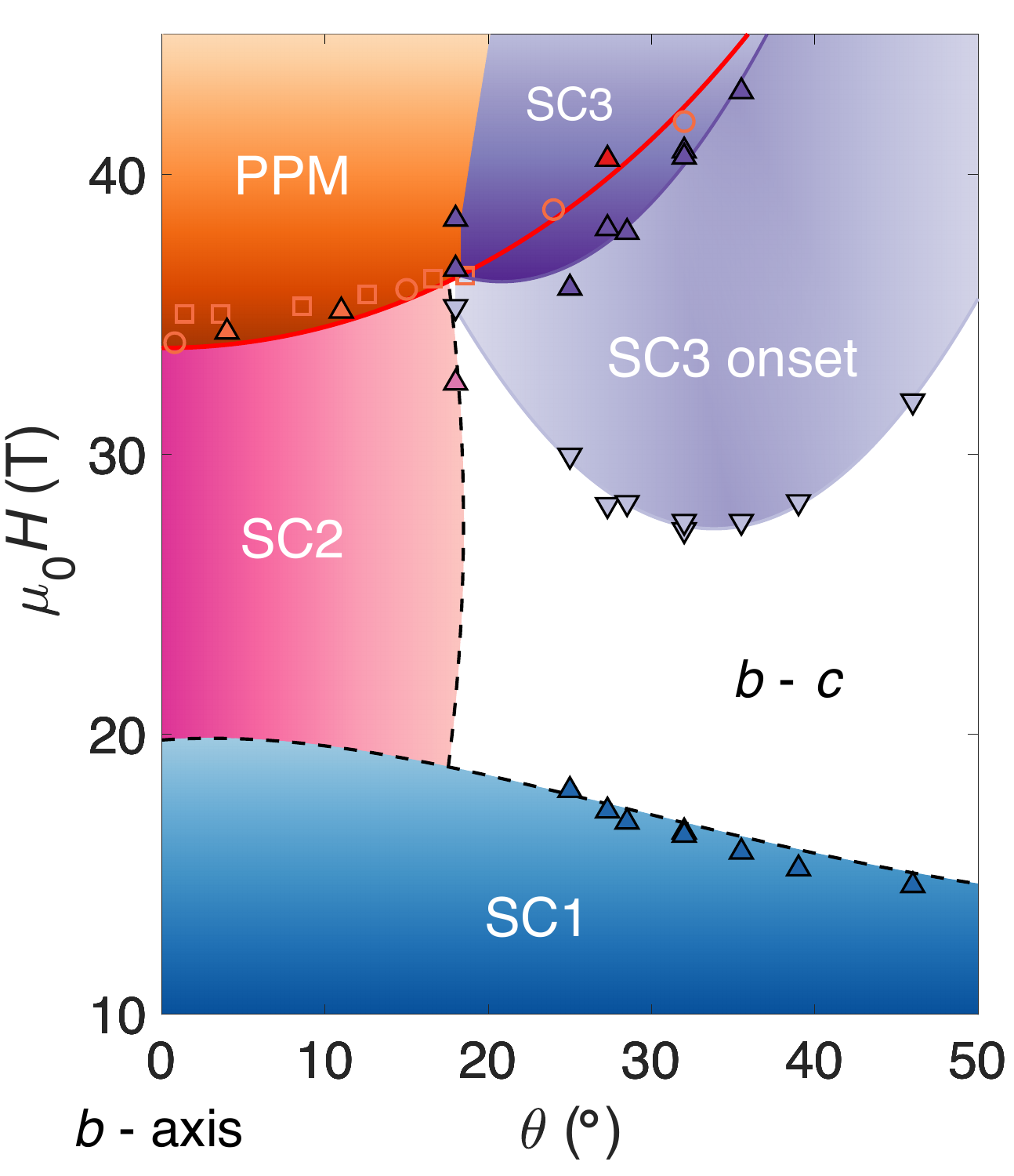}
    \caption{UTe$_2$ low-$T$ high-$H$ phase diagram for the $b-c$ rotation plane. All triangular data points are from this study. Square points and the red fit line for $H^*(\theta)$ are reproduced from \cite{tony2024enhanced,pnas24data}, while circular points are from \cite{qcl,qcldata}. PPM stands for polarized paramagnet. The onset of SC3 is determined by the maximum in $\rho(H)$ of the curves in Fig.~\ref{fig:angles}, while the SC3 region itself is defined by the observation of zero resistance.}
    \label{fig:phasediag}
\end{figure}

We summarize our results in Fig.~\ref{fig:phasediag}, presenting an updated $H-\theta$ UTe$_2$ phase diagram for the $b-c$ rotation plane in the low temperature limit. We use upwards triangular points to demarcate phase boundaries (for example, identified by transitions between $\rho=0$ and $\rho \neq 0$), while downwards triangles mark the locations of maxima in $\rho(H)$, which appear to indicate the onset of SC3~\cite{qcl}. This onset region occupies a broad portion of the $H-\theta$ phase space, as previously observed in the $b-ac$ rotation plane~\cite{qcl}.

Orange triangles mark the first-order phase boundary into the spin-polarized state, as identified by the sudden increase in $\rho$ observed at 4$\degree$ and 11$\degree$ in Fig.~\ref{fig:angles} for $T = 0.4$~K. The red triangle represents the location of $H^*$ discerned by the 8~K measurement in Fig.~\ref{fig:temps}, while the open orange symbols and red line are reproduced from~\cite{tony2024enhanced,pnas24data,qcl,qcldata} that reported prior measurements of $H^*(\theta)$. The key finding of this study is that there is a small region of the $H-\theta$ phase space in which the SC3 state spills out beyond the polarized paramagnetic state, as depicted by the upwards purple triangles lying at lower values of $H$ (for the same $\theta$) than the metamagnetic phase boundary (red line).

\section{Discussion}

A number of independent groups have investigated the physical properties of the SC3 state~\cite{Ranfieldboostednatphys2019,knafo2021comparison,ran2021expansion,lewin2023review,Miyake2021,LANL_bulk_UTe2,helm2024,qcl,lewin2025halo,tony2025brief,moir2025pnas,marquardt2025bulksignaturesreentrantsuperconductivity}. One notable recent discovery was the observation of SC3 superconductivity at high fields in samples with very high levels of disorder, sufficiently dirty so as to inhibit the formation of the SC1 state in zero field~\cite{frank2024orphan}. Such immunity to disorder could be taken to suggest that SC3 is formed by a conventional pairing method, which would fit with an interpretation of a field-compensation mechanism~\cite{JaccPet1962PRL} overcoming orbital and Pauli pair-breaking effects to enable a conventional singlet superconductive state to exist at high $H$~\cite{helm2024}.

However, a closer inspection of how the properties of the SC3 state depend on disorder content provides strong evidence in favor of unconventional pairing. It is well known that unconventional superconductors, in which the pair wave function possesses angular momentum $l > 0$ and the superconducting order parameter is of lower symmetry than the underlying lattice, exhibit a pronounced sensitivity to the presence of non-magnetic impurities~\cite{APMackenzie_SRO_RRR}. The effects of disorder become considerable when the ratio between the coherence length and mean free path is close to unity, with the coherence length given by $\xi_0 = \sqrt{\nicefrac{\phi_0}{2\pi B_{c2}}}$. We expect $\nicefrac{T_c}{T_{c0}}$ to vary approximately as $\nicefrac{\rho_0}{\sqrt{B_{c2}}}$ \cite{tilley_superbook}, with $T_{c0}$ the maximal critical temperature, $\rho_0$ the residual resistivity indicative of disorder content and $B_{c2}$ the upper critical field. For unconventional pairing, samples with greater disorder content should therefore exhibit lower $T_c$ and $B_{c2}$ values. This is indeed what has been observed for SC3 in UTe$_2$. In \cite{tony2025brief} measurements on a high quality specimen report a maximal $T_c$ of $\approx$~2.4~K, $B_{c2} \gtrapprox$~70~T and a wide angular range of SC3 in the $bc$ plane spanning $\approx 20\degree$ at 0.7~K. By contrast, in \cite{frank2024orphan} the lowest quality sample exhibited $T_c \approx 0.8$~K, $B_{c2} \lessapprox 55$~T over a span of $\approx 12\degree$ at 0.5~K. Samples of intermediate quality fit within these ranges~\cite{frank2024orphan,tony2025brief,qcl,helm2024,Ranfieldboostednatphys2019,lewin2023review}, with higher $T_c$ generally correlating with higher $B_{c2}$ and a wider angular span. We conclude that the survival of superconductivity in the given impure sample \cite{frank2024orphan} is not inconsistent with unconventional pairing -- on the contrary, the collapse in size of the superconducting regime in field provides significant support for unconventional pairing.

Recently, a doping-dependence study of UTe$_2$ was conducted~\cite{moir2025pnas}, which examined the effect of substituting uranium for thorium. In contrast to \cite{frank2024orphan}, in \cite{moir2025pnas} the SC3 phase was found to be absent for thorium substitution $\geq$2.5\%, whereas SC1 persisted up to the maximally-studied doping of 4.7\%. This indicates that the different superconducting phases of UTe$_2$ have contrasting sensitivities to the specific type of disorder content present in a given sample. Stoichiometric UTe$_2$ crystallizes such that the uranium atoms form a ladder configuration extending along the [100] direction. The shortest U-U distance is along the rungs of these ladders, where the uraniums are separated by 3.72~\AA~in the [001] direction~\cite{Hutanu}. As thorium atoms will selectively occupy uranium sites, the results of \cite{moir2025pnas} suggest a central role of the uranium-uranium dimer configuration in UTe$_2$ governing the manifestation of the SC3 and spin-polarized phases. The presence of ferromagnetic fluctuations along this short dimer distance has previously been discussed in the context of neutron scattering measurements~\cite{Knafo104.L100409}. It is worth noting that the quasi-2D electronic structure~\cite{AokidHvA_UTe2-2022,Eaton2024,theo2024,zhang2025dimensionality,weinberger2025pressure} may also be important for the stabilization of superconductivity at such high $H$, as is common in other systems~\cite{Norman_review_doi:10.1126/science.1200181,Stewart_review_doi:10.1080/00018732.2017.1331615}. If pairing of the SC3 state is indeed mediated by low-$Q$ ferromgneticlike fluctuations, that would point strongly toward an unconventional pseudospin-triplet state, potentially of non-unitary character~\cite{lewin2025halo}.

In our measurements here, we identified a substantial portion of $H-\theta$ phase space to be occupied by the onset of the SC3 phase, identified by an anomalous negative gradient of $\rho(H)$ before $\rho$ falls to zero at higher $H$ as SC3 fully condenses. This onset region is broadest -- that is, it extends to lowest $H$ -- at $\theta \approx 34\degree$ (Fig.~\ref{fig:phasediag}). Recent pulsed-field magnetotransport measurements of UTe$_2$ have found that upon cooling the SC3 phase condenses out of a strange metallic `normal' state, identified by the presence of linear-in-temperature resistivity in the temperature range immediately above $T_c$~\cite{weinberger2025strangemetallicityencompasseshigh}, which rises at a rate indicative of Planckian dissipation~\cite{Hartnoll+AndyRevModPhys.94.041002}. Interestingly, in the strange metal state the resistivity $A$ coefficient is sharply peaked around 34$\degree$, coinciding with the greatest extent of the SC3 onset region we identify here. This is also where the upper critical field and critical temperature of SC3 are highest~\cite{tony2025brief}. In combination, these observations strongly point to the SC3 state emerging from a quantum critical regime.

However, although the high-$H$ metamagnetic transition surface of UTe$_2$ has been shown to terminate at a quantum critical phase boundary~\cite{qcl}, this occurs comparatively far from $\theta = 34\degree$ where both SC3 and strange metallicity are most pronounced. It therefore remains an outstanding puzzle as to what order parameter might be attaining criticality in this region of the complex phase diagram. Preliminary observations of anomalous magnetotransport signatures~\cite{LewinPRB24,ripples} suggest that sub-leading magnetic phases may coexist either within or close to the spin-polarized state, which may be analogous to the cases of some other metamagnets~\cite{knafo2016field,lester2015field,lester2021magnetic}. Further experimental and theoretical efforts are required to better understand the precise microscopic details underpinning the formation of the enigmatic SC3 state~\cite{YuPauli26}.

In summary, we performed magnetotransport measurements of a high quality UTe$_2$ single crystal at low temperatures and applied steady magnetic field strengths up to 45~T, tilted in the crystallographic $bc$ plane. We identify a small region of the $H-\theta$ phase diagram in which the high-$H$ SC3 superconductive state condenses outside the spin-polarized region, as identified by the presence of zero resistivity. We also observe a broad onset region of the SC3 state, extending down to fields as low as 27.3~T. This onset region is greatest in the vicinity that pulsed-field magnetotransport measurements at higher $H$ and $T$ have observed strange metallic linear-$T$ resistivity~\cite{weinberger2025strangemetallicityencompasseshigh} -- which also coincides with where the SC3 state exhibits its greatest $T_c$ and $B_{c2}$~\cite{tony2025brief}. In combination, these observations strongly point to a quantum critical origin of the SC3 state of UTe$_2$.

\vspace{-0mm}
\begin{acknowledgments}\vspace{-5mm}
We gratefully acknowledge stimulating discussions with D. Chichinadze, D. Shaffer, J. Yu and S. Raghu. We thank G. Knebel for astute feedback. This project was supported by the EPSRC of the UK (grants EP/Z533695/1, EP/X011992/1 and EP/Z533695/1). Crystal growth and characterization were performed in MGML (mgml.eu), which is supported within the program of Czech Research Infrastructures (project no. LM2023065). We acknowledge financial support by the Czech Science Foundation (GACR), project No. 22-22322S.  A portion of this work was performed at the National High Magnetic Field Laboratory, which is supported by National Science Foundation Cooperative Agreement No. DMR-2128556 and the State of Florida. T.I.W. and A.G.E. acknowledge support from ICAM through US National Science Foundation (NSF) Grant Number 2201516 under the Accelnet program of the Office of International Science and Engineering and from QuantEmX grants from ICAM and the Gordon and Betty Moore Foundation through Grant GBMF9616. A.G.E. acknowledges support from Sidney Sussex College (University of Cambridge).
\end{acknowledgments}

%\clearpage
\bibliography{UTe2}

@misc{ripples,
      title={{Metamagnetic ripples in the UTe$_2$ high magnetic field phase diagram}}, 
      author={Zheyu Wu and Hanyi Chen and Mengmeng Long and Gangjian Jin and Huakun Zuo and Daniel Shaffer and Dmitry V. Chichinadze and Andrej Cabala and Vladimir Sechovsky and Michal Valiska and Zengwei Zhu and Gilbert G. Lonzarich and F. Malte Grosche and Alexander G. Eaton},
      year={2025},
      eprint={2503.11362},
      archivePrefix={arXiv}
}

@article{moir2025pnas,
  title={{High-magnetic-field phases in U$_{1-x}$Th$_x$Te$_2$}},
  author={Moir, Camilla M and Singleton, John and Blawat, Joanna and Lee-Wong, Eric and Deng, Yuhang and Feng, Keke and Wannamaker, Tyler and Baumbach, Ryan E and Maple, M Brian},
  journal={Proc. Natl. Acad. Sci. (USA)},
  volume={122},
  number={42},
  pages={e2521261122},
  year={2025},
  publisher={National Academy of Sciences},
  url={https://doi.org/10.1073/pnas.2521261122}
}

@misc{weinberger2025strangemetallicityencompasseshigh,
      title={{Strange metallicity encompasses high magnetic field-induced superconductivity in UTe$_2$}}, 
      author={T. I. Weinberger and H. Chen and Z. Wu and M. Long and A. Cabala and Y. Skourski and J. Sourd and T. Haidamak and V. Sechovsky and M. Valiska and F. M. Grosche and A. G. Eaton},
      year={2025},
      eprint={2505.12131},
      archivePrefix={arXiv}
}

@article{LewinPRB24,
  title = {{Field-angle evolution of the superconducting and magnetic phases of ${\mathrm{UTe}}_{2}$ around the $b$ axis}},
  author = {Lewin, Sylvia K. and Yu, Josephine J. and Frank, Corey E. and Graf, David and Chen, Patrick and Ran, Sheng and Eo, Yun Suk and Paglione, Johnpierre and Raghu, S. and Butch, Nicholas P.},
  journal = {Phys. Rev. B},
  volume = {110},
  issue = {18},
  pages = {184520},
  numpages = {16},
  year = {2024},
  month = {Nov},
  publisher = {American Physical Society},
  doi = {10.1103/PhysRevB.110.184520},
  url = {https://link.aps.org/doi/10.1103/PhysRevB.110.184520}
}

@article{YuPauli26,
  title = {{Pauli ``unlimited'': Magnetic field induced superconductivity in ${\mathrm{UTe}}_{2}$}},
  author = {Yu, Josephine J. and Yu, Yue and Murthy, Chaitanya and Raghu, S.},
  journal = {Phys. Rev. B},
  volume = {113},
  issue = {1},
  pages = {014517},
  numpages = {12},
  year = {2026},
  month = {Jan},
  publisher = {American Physical Society},
  doi = {10.1103/3ldv-ndqj},
  url = {https://link.aps.org/doi/10.1103/3ldv-ndqj}
}

@article{Stewart_review_doi:10.1080/00018732.2017.1331615,
author = {G. R. Stewart},
title = {Unconventional superconductivity},
journal = {Adv. Phys.},
volume = {66},
number = {2},
pages = {75-196},
year  = {2017},
publisher = {Taylor & Francis},
doi = {10.1080/00018732.2017.1331615},

URL = { 
    
        https://doi.org/10.1080/00018732.2017.1331615
},


}

@article{
Norman_review_doi:10.1126/science.1200181,
author = {Michael R. Norman },
title = {{The Challenge of Unconventional Superconductivity}},
journal = {Science},
volume = {332},
number = {6026},
pages = {196-200},
year = {2011},
doi = {10.1126/science.1200181},
URL = {https://www.science.org/doi/abs/10.1126/science.1200181}}

@article{helm2024,
  title={{Field-induced compensation of magnetic exchange as the possible origin of reentrant superconductivity in UTe$_2$}},
  author={Helm, Toni and Kimata, Motoi and Sudo, Kenta and Miyata, Atsuhiko and Stirnat, Julia and F{\"o}rster, Tobias and Hornung, Jacob and K{\"o}nig, Markus and Sheikin, Ilya and Pourret, Alexandre and others},
  journal={Nat. Commun.},
  volume={15},
  number={1},
  pages={37},
  year={2024},
  url={https://doi.org/10.1038/s41467-023-44183-1}
}

@article{LANL_bulk_UTe2,
  title={{Sudden adiabaticity signals reentrant bulk superconductivity in UTe$_2$}},
  author={Sch{\"o}nemann, Rico and Rosa, Priscila FS and Thomas, Sean M and Lai, You and Nguyen, Doan N and Singleton, John and Brosha, Eric L and McDonald, Ross D and Zapf, Vivien and Maiorov, Boris and others},
  journal={PNAS Nexus},
  volume={3},
  number={1},
  pages={pgad428},
  year={2024},
  url={https://doi.org/10.1093/pnasnexus/pgad428}
}

@article{qcl,
  title = {{A Quantum Critical Line Bounds the High Field Metamagnetic Transition Surface in ${\mathrm{UTe}}_{2}$}},
  author = {Wu, Z. and Weinberger, T. I. and Hickey, A. J. and Chichinadze, D. V. and Shaffer, D. and Cabala, A. and Chen, H. and Long, M. and Brumm, T. J. and Xie, W. and Ling, Y. and Zhu, Z. and Skourski, Y. and Graf, D. E. and Sechovsk\'y, V. and Vali\ifmmode \check{s}\else \v{s}\fi{}ka, M. and Lonzarich, G. G. and Grosche, F. M. and Eaton, A. G.},
  journal = {Phys. Rev. X},
  volume = {15},
  issue = {2},
  pages = {021019},
  numpages = {20},
  year = {2025},
  month = {Apr},
  publisher = {American Physical Society},
  doi = {10.1103/PhysRevX.15.021019},
  url = {https://link.aps.org/doi/10.1103/PhysRevX.15.021019}
}

@misc{pnas24data,
      title={{Research data supporting: Enhanced triplet superconductivity in next generation ultraclean UTe$_2$}}, 
      author={Z. Wu  and T. I. Weinberger  and J. Chen  and A. Cabala  and D. V. Chichinadze  and D. Shaffer  and J. Pospíšil  and J. Prokleška  and T. Haidamak  and G. Bastien  and V. Sechovský  and A. J. Hickey  and M. J. Mancera-Ugarte  and S. Benjamin  and D. E. Graf  and Y. Skourski  and G. G. Lonzarich  and M. Vališka  and F. M. Grosche  and A. G. Eaton},
      year={2024},
      note={{University of Cambridge Apollo Repository}},
      url={http://doi.org/10.17863/CAM.110675}, 
}

@misc{qcldata,
      title={{Research data supporting: A quantum critical line bounds the high field metamagnetic transition surface in UTe$_2$}}, 
      author={Wu, Z. and Weinberger, T. I. and Hickey, A. J. and Chichinadze, D. V. and Shaffer, D. and Cabala, A. and Chen, H. and Long, M. and Brumm, T. J. and Xie, W. and Ling, Y. and Zhu, Z. and Skourski, Y. and Graf, D. E. and Sechovsk\'y, V. and Vali\ifmmode \check{s}\else \v{s}\fi{}ka, M. and Lonzarich, G. G. and Grosche, F. M. and Eaton, A. G.},
      year={2025},
      note={{University of Cambridge Apollo Repository}},
      url={http://doi.org/10.17863/CAM.116724}, 
}

@article{tony2024enhanced,
  title={{Enhanced triplet superconductivity in next-generation ultraclean UTe$_2$}},
  author = {Z. Wu  and T. I. Weinberger  and J. Chen  and A. Cabala  and D. V. Chichinadze  and D. Shaffer  and J. Pospíšil  and J. Prokleška  and T. Haidamak  and G. Bastien  and V. Sechovský  and A. J. Hickey  and M. J. Mancera-Ugarte  and S. Benjamin  and D. E. Graf  and Y. Skourski  and G. G. Lonzarich  and M. Vališka  and F. M. Grosche  and A. G. Eaton},
  journal={Proc. Natl. Acad. Sci. USA},
  volume={121},
  number={37},
  pages={e2403067121},
  year={2024},
  publisher={National Academy of Sciences},
  url={https://doi.org/10.1073/pnas.2403067121}
}

@article{tony2025brief,
  title={{Superconducting critical temperature elevated by intense magnetic fields}},
  author = {Z. Wu  and H. Chen and T. I. Weinberger  and A. Cabala and D. E. Graf  and Y. Skourski and W. Xie and Y. Ling and Z. Zhu and V. Sechovský  and M. Vališka  and F. M. Grosche  and A. G. Eaton},
  journal={Proc. Natl. Acad. Sci. USA},
  volume={122},
  number={2},
  pages={e2422156122},
  year={2025},
  publisher={National Academy of Sciences},
  url={https://doi.org/10.1073/pnas.2422156122}
}

@article{JaccPet1962PRL,
  title = {{Ultra-High-Field Superconductivity}},
  author = {Jaccarino, V. and Peter, M.},
  journal = {Phys. Rev. Lett.},
  volume = {9},
  issue = {7},
  pages = {290--292},
  numpages = {0},
  year = {1962},
  month = {Oct},
  publisher = {American Physical Society},
  doi = {10.1103/PhysRevLett.9.290},
  url = {https://link.aps.org/doi/10.1103/PhysRevLett.9.290}
}

@article{knafo2019magnetic,
  title={{Magnetic-field-induced phenomena in the paramagnetic superconductor ${\mathrm{UTe}}_{2}$}},
  author={Knafo, William and Vali{\v{s}}ka, Michal and Braithwaite, Daniel and Lapertot, G{\'e}rard and Knebel, Georg and Pourret, Alexandre and Brison, Jean-Pascal and Flouquet, Jacques and Aoki, Dai},
  journal={J. Phys. Soc. Jpn.},
  volume={88},
  number={6},
  pages={063705},
  year={2019},
  publisher={The Physical Society of Japan},
  url = {https://journals.jps.jp/doi/suppl/10.7566/JPSJ.88.063705}
}

@article{frank2024orphan,
  title={Orphan high field superconductivity in non-superconducting uranium ditelluride},
  author={Frank, Corey E and Lewin, Sylvia K and Saucedo Salas, Gicela and Czajka, Peter and Hayes, Ian M and Yoon, Hyeok and Metz, Tristin and Paglione, Johnpierre and Singleton, John and Butch, Nicholas P},
  journal={Nat. Commun.},
  volume={15},
  pages={3378},
  year={2024},
  publisher={Nature Publishing Group UK London},
  url={https://doi.org/10.1038/s41467-024-47090-1}
}

@book{tilley_superbook,
  title={Superfluidity and superconductivity},
  author={Tilley, David R and Tilley, John},
  year={1990},
  publisher={Routledge, Oxfordshire, UK},
  url={https://doi.org/10.1201/9780203737897}
}

@article{valiska2024dramatic,
  title = {{Dramatic elastic response at the critical end point in ${\mathrm{UTe}}_{2}$}},
  author = {Michal Vališka and Tetiana Haidamak and Andrej Cabala and Jiří Pospíšil and Gaël Bastien and Vladimír Sechovský and Jan Prokleška and Tatsuya Yanagisawa and Petr Opletal and Hironori Sakai and Yoshinori Haga and Atsuhiko Miyata and Denis Gorbunov and Sergei Zherlitsyn},
  journal = {Phys. Rev. Mater.},
  volume = {8},
  issue = {9},
  pages = {094415},
  numpages = {9},
  year = {2024},
  month = {Sep},
  publisher = {American Physical Society},
  doi = {10.1103/PhysRevMaterials.8.094415},
  url = {https://link.aps.org/doi/10.1103/PhysRevMaterials.8.094415}
}

@article{ran2021expansion,
  title={{Expansion of the high field-boosted superconductivity in UTe$_2$ under pressure}},
  author={Ran, Sheng and Saha, Shanta R and Liu, I-Lin and Graf, David and Paglione, Johnpierre and Butch, Nicholas P},
  journal={npj Quantum Mater.},
  volume={6},
  number={1},
  pages={75},
  year={2021},
  publisher={Nature Publishing Group UK London},
  url={https://doi.org/10.1038/s41535-021-00376-9}
}

@article{Knafo104.L100409,
  title = {{Low-dimensional antiferromagnetic fluctuations in the heavy-fermion paramagnetic ladder compound ${\mathrm{UTe}}_{2}$}},
  author = {Knafo, W. and Knebel, G. and Steffens, P. and Kaneko, K. and Rosuel, A. and Brison, J.-P. and Flouquet, J. and Aoki, D. and Lapertot, G. and Raymond, S.},
  journal = {Phys. Rev. B},
  volume = {104},
  issue = {10},
  pages = {L100409},
  numpages = {6},
  year = {2021},
  month = {Sep},
  publisher = {American Physical Society},
  doi = {10.1103/PhysRevB.104.L100409},
  url = {https://link.aps.org/doi/10.1103/PhysRevB.104.L100409}
}

@article{knafo2021comparison,
  title={{Comparison of two superconducting phases induced by a magnetic field in UTe$_2$}},
  author={Knafo, William and Nardone, M and Vali{\v{s}}ka, M and Zitouni, A and Lapertot, G and Aoki, D and Knebel, G and Braithwaite, D},
  journal={Commun. Phys.},
  volume={4},
  number={1},
  pages={40},
  year={2021},
  publisher={Nature Publishing Group},
  url = {https://doi.org/10.1038/s42005-021-00545-z},
}

@article{lewin2023review,
  title={{A Review of UTe$_2$ at High Magnetic Fields}},
  author={Lewin, Sylvia K and Frank, Corey E and Ran, Sheng and Paglione, Johnpierre and Butch, Nicholas P},
  journal={Rep. Prog. Phys.},
  year={2023},
  volume={86},
  number={11},
  pages={114501},
  publisher={IOP Publishing},
  url={https://doi.org/10.1088/1361-6633/acfb93}
}

@article{weinberger2025pressure,
  title={{Pressure-enhanced $f$-electron orbital weighting in UTe$_2$ mapped by quantum interferometry}},
  author={Weinberger, TI and Wu, Z and Hickey, AJ and Graf, DE and Li, G and Wang, P and Zhou, R and Cabala, A and Pu, J and Sechovsky, V and Valiska, M and Lonzarich, GG and Grosche, FM and Eaton, AG},
  journal={Commun. Phys.},
  volume={8},
  number={1},
  pages={454},
  year={2025},
  publisher={Nature Publishing Group UK London},
  url={https://doi.org/10.1038/s42005-025-02333-5}
}

@article{zhang2025dimensionality,
  title={{Dimensionality of the reinforced superconductivity in UTe$_2$}},
  author={Zhang, L and Guo, C and Graf, D and Putzke, C and Bordelon, MM and Bauer, ED and Thomas, SM and Ronning, F and Rosa, PFS and Moll, PJW},
  journal={Nat. Commun.},
  volume={16},
  number={1},
  pages={10308},
  year={2025},
  publisher={Nature Publishing Group UK London},
  url={https://doi.org/10.1038/s41467-025-66288-5}
}

@misc{marquardt2025bulksignaturesreentrantsuperconductivity,
      title={Bulk signatures of re-entrant superconductivity in UTe$_2$ from ultrasound measurements}, 
      author={N. Marquardt and C. Duffy and C. Proust and S. Badoux and M. Amano Patino and G. Lapertot and D. Aoki and J. -P. Brison and G. Knebel and D. LeBoeuf},
      year={2025},
      eprint={2512.17691},
      archivePrefix={arXiv} 
}

@article{Hartnoll+AndyRevModPhys.94.041002,
  title = {{Colloquium: Planckian dissipation in metals}},
  author = {Hartnoll, Sean A. and Mackenzie, Andrew P.},
  journal = {Rev. Mod. Phys.},
  volume = {94},
  issue = {4},
  pages = {041002},
  numpages = {27},
  year = {2022},
  month = {Nov},
  publisher = {American Physical Society},
  doi = {10.1103/RevModPhys.94.041002},
  url = {https://link.aps.org/doi/10.1103/RevModPhys.94.041002}
}

@article{Aoki_UTe2review2022,
author = {Aoki, D. and Brison, J. P. and Flouquet, J. and Ishida, K. and Knebel, G. and Tokunaga, Y. and Yanase, Y.},
doi = {10.1088/1361-648X/AC5863},
issn = {0953-8984},
journal = {J. Phys. Condens. Matter},
month = {apr},
number = {24},
pages = {243002},
pmid = {35203074},
publisher = {IOP Publishing},
title = {{Unconventional superconductivity in UTe$_2$}},
url = {https://iopscience.iop.org/article/10.1088/1361-648X/ac5863 https://iopscience.iop.org/article/10.1088/1361-648X/ac5863/meta},
volume = {34},
year = {2022}
}

@article{Ranfieldboostednatphys2019,
author = {Ran, Sheng and Liu, I. Lin and Eo, Yun Suk and Campbell, Daniel J. and Neves, Paul M. and Fuhrman, Wesley T. and Saha, Shanta R. and Eckberg, Christopher and Kim, Hyunsoo and Graf, David and Balakirev, Fedor and Singleton, John and Paglione, Johnpierre and Butch, Nicholas P.},
doi = {10.1038/s41567-019-0670-x},
issn = {1745-2481},
journal = {Nat. Phys.},
month = {oct},
number = {12},
pages = {1250--1254},
publisher = {Nature Publishing Group},
title = {{Extreme magnetic field-boosted superconductivity}},
url = {https://www.nature.com/articles/s41567-019-0670-x},
volume = {15},
year = {2019}
}

@article{Ran2019Science,
author = {Ran, Sheng and Eckberg, Chris and Ding, Qing Ping and Furukawa, Yuji and Metz, Tristin and Saha, Shanta R. and Liu, I. Lin and Zic, Mark and Kim, Hyunsoo and Paglione, Johnpierre and Butch, Nicholas P.},
doi = {10.1126/SCIENCE.AAV8645/SUPPL_FILE/AAV8645_RAN_SM.PDF},
issn = {10959203},
journal = {Science},
month = {aug},
number = {6454},
pages = {684--687},
pmid = {31416960},
publisher = {American Association for the Advancement of Science},
title = {{Nearly ferromagnetic spin-triplet superconductivity}},
url = {https://www.science.org/doi/10.1126/science.aav8645},
volume = {365},
year = {2019}
}

@article{Miyake2019,
author = {Miyake, Atsushi and Shimizu, Yusei and Sato, Yoshiki J. and Li, Dexin and Nakamura, Ai and Homma, Yoshiya and Honda, Fuminori and Flouquet, Jacques and Tokunaga, Masashi and Aoki, Dai},
issn = {13474073},
journal = {J. Phys. Soc. Jpn.},
month = {may},
number = {6},
publisher = {The Physical Society of Japan},
title = {{Metamagnetic Transition in Heavy Fermion Superconductor ${\mathrm{UTe}}_{2}$}},
url = {https://journals.jps.jp/doi/abs/10.7566/JPSJ.88.063706},
volume = {88},
year = {2019}
}

@article{Knebel2019,

author = {Knebel, Georg and Knafo, William and Pourret, Alexandre and Niu, Qun and Vali{\v{s}}ka, Michal and Braithwaite, Daniel and Lapertot, G{\'{e}}rard and Nardone, Marc and Zitouni, Abdelaziz and Mishra, Sanu and Sheikin, Ilya and Seyfarth, Gabriel and Brison, Jean Pascal and Aoki, Dai and Flouquet, Jacques},
doi = {10.7566/JPSJ.88.063707},
issn = {13474073},
journal = {J. Phys. Soc. Jpn.},
month = {may},
number = {6},
pages = {63707},
publisher = {The Physical Society of Japan},
title = {{Field-Reentrant Superconductivity Close to a Metamagnetic Transition in the Heavy-Fermion Superconductor ${\mathrm{UTe}}_{2}$}},
url = {https://journals.jps.jp/doi/abs/10.7566/JPSJ.88.063707},
volume = {88},
year = {2019}
}

@article{AokidHvA_UTe2-2022,
author = {Aoki, Dai and Hironori, Sakai and Petr, Opletal and Yoshifumi, Tokiwa and Jun, Ishizuka and Youichi, Yanase and Hisatomo, Harima and Ai, Nakamura and Dexin, Li and Yoshiya, Homma and Yusei, Shimizu and Georg, Knebel and Jacques, Flouquet and Yoshinori, Haga},
doi = {10.7566/JPSJ.91.083704},
issn = {0031-9015},
journal = {J. Phys. Soc. Jpn.},
month = {jul},
number = {8},
pages={083704},
publisher = {The Physical Society of Japan},
title = {{First Observation of the de Haas–van Alphen Effect and Fermi Surfaces in the Unconventional Superconductor UTe$_2$}},
url = {https://journals.jps.jp/doi/abs/10.7566/JPSJ.91.083704},
volume = {91},
year = {2022}
}

@article{Miyake2021,
author = {Miyake, Atsushi and Shimizu, Yusei and Sato, Yoshiki J. and Li, Dexin and Nakamura, Ai and Homma, Yoshiya and Honda, Fuminori and Flouquet, Jacques and Tokunaga, Masashi and Aoki, Dai},
issn = {13474073},
journal = {J. Phys. Soc. Jpn.},
month = {sep},
number = {10},
pages = {103702},
publisher = {The Physical Society of Japan},
title = {{Enhancement and Discontinuity of Effective Mass through the First-Order Metamagnetic Transition in ${\mathrm{UTe}}_{2}$}},
url = {https://journals.jps.jp/doi/abs/10.7566/JPSJ.90.103702},
volume = {90},
year = {2021}
}

@article{lewin2025halo,
  title={{High-field superconducting halo in UTe$_2$}},
  author={Lewin, Sylvia K and Czajka, Peter and Frank, Corey E and Saucedo Salas, Gicela and Noe II, G Timothy and Yoon, Hyeok and Eo, Yun Suk and Paglione, Johnpierre and Nevidomskyy, Andriy H and Singleton, John and Butch, Nicholas P},
  journal={Science},
  volume={389},
  number={6759},
  pages={512--515},
  year={2025},
  publisher={American Association for the Advancement of Science},
  url={https://doi.org/10.1126/science.adn7673}
}

@article{Hutanu,
author = "Hutanu, V. and Deng, H. and Ran, S. and Fuhrman, W. T. and Thoma, H. and Butch, N. P.",
title = "{Low-temperature crystal structure of the unconventional spin-triplet superconductor {UTe$_2$} from single-crystal neutron diffraction}",
journal = " Acta Crystallogr. B",
year = "2020",
volume = "76",
number = "1",
pages = "137--143",
month = "Feb",
doi = {10.1107/S2052520619016950},
url = {https://doi.org/10.1107/S2052520619016950},

}

@article{Eaton2024,
   author = {A. G. Eaton and T. I. Weinberger and N. J. M. Popiel and Z. Wu and A. J. Hickey and A. Cabala and J. Pospíšil and J. Prokleška and T. Haidamak and G. Bastien and P. Opletal and H. Sakai and Y. Haga and R. Nowell and S. M. Benjamin and V. Sechovský and G. G. Lonzarich and F. M. Grosche and M. Vališka},
   doi = {10.1038/s41467-023-44110-4},
   journal = {Nat. Commun.},
   pages = {223},
   publisher = {Nature Publishing Group},
   title = {{Quasi-2D Fermi surface in the anomalous superconductor UTe$_2$}},
   volume = {15},
   url = {https://doi.org/10.1038/s41467-023-44110-4},
   year = {2024}
}

@article{APMackenzie_SRO_RRR,
  title = {{Extremely Strong Dependence of Superconductivity on Disorder in ${\mathrm{Sr}}_{2}{\mathrm{RuO}}_{4}$}},
  author = {Mackenzie, A. P. and Haselwimmer, R. K. W. and Tyler, A. W. and Lonzarich, G. G. and Mori, Y. and Nishizaki, S. and Maeno, Y.},
  journal = {Phys. Rev. Lett.},
  volume = {80},
  issue = {1},
  pages = {161--164},
  numpages = {0},
  year = {1998},
  month = {Jan},
  publisher = {American Physical Society},
  doi = {10.1103/PhysRevLett.80.161},
  url = {https://link.aps.org/doi/10.1103/PhysRevLett.80.161}
}

@article{lester2021magnetic,
  title={{Magnetic-field-controlled spin fluctuations and quantum criticality in Sr$_3$Ru$_2$O$_7$}},
  author={Lester, C and Ramos, S and Perry, RS and Croft, TP and Laver, M and Bewley, RI and Guidi, T and Hiess, A and Wildes, A and Forgan, EM and others},
  journal={Nat. Commun.},
  volume={12},
  number={1},
  pages={5798},
  year={2021},
  publisher={Nature Publishing Group UK London},
  url={https://doi.org/10.1038/s41467-021-26068-3}
}

@article{lester2015field,
  title={{Field-tunable spin-density-wave phases in Sr$_3$Ru$_2$O$_7$}},
  author={Lester, C and Ramos, Silvia and Perry, RS and Croft, TP and Bewley, RI and Guidi, T and Manuel, P and Khalyavin, DD and Forgan, EM and Hayden, SM},
  journal={Nat. Mater.},
  volume={14},
  number={4},
  pages={373--378},
  year={2015},
  publisher={Nature Publishing Group UK London},
  url={https://doi.org/10.1038/nmat4181}
}

@article{knafo2016field,
  title={{Field-induced spin-density wave beyond hidden order in URu$_2$Si$_2$}},
  author={Knafo, William and Duc, F and Bourdarot, Frederic and Kuwahara, K and Nojiri, H and Aoki, D and Billette, Julien and Frings, P and Tonon, X and Leli{\`e}vre-Berna, E and others},
  journal={Nat. Commun.},
  volume={7},
  number={1},
  pages={13075},
  year={2016},
  publisher={Nature Publishing Group UK London},
  url={https://doi.org/10.1038/ncomms13075}
}

@article{theo2024,
  title = {{Quantum Interference between Quasi-2D Fermi Surface Sheets in ${\mathrm{UTe}}_{2}$}},
  author = {Weinberger, T. I. and Wu, Z. and Graf, D. E. and Skourski, Y. and Cabala, A. and Posp\'{\i}\ifmmode \check{s}\else \v{s}\fi{}il, J. and Prokle\ifmmode \check{s}\else \v{s}\fi{}ka, J. and Haidamak, T. and Bastien, G. and Sechovsk\'y, V. and Lonzarich, G. G. and Vali\ifmmode \check{s}\else \v{s}\fi{}ka, M. and Grosche, F. M. and Eaton, A. G.},
  journal = {Phys. Rev. Lett.},
  volume = {132},
  issue = {26},
  pages = {266503},
  numpages = {8},
  year = {2024},
  month = {Jun},
  publisher = {American Physical Society},
  doi = {10.1103/PhysRevLett.132.266503},
  url = {https://link.aps.org/doi/10.1103/PhysRevLett.132.266503}
}
\end{document}